\documentclass[reprint,aps,prb,superscriptaddress]{revtex4-1}
\pdfoutput=1

\usepackage{bm}
\usepackage{amsmath,amsthm,amssymb}
\usepackage{array,booktabs,longtable}
\usepackage{mathrsfs}
\usepackage{subfigure}
\usepackage{color}
\usepackage{tikz}
\usetikzlibrary{decorations.markings}
\usetikzlibrary{calc,shapes,positioning,arrows,snakes}
\usepackage{array}
\usepackage{verbatim}
\usepackage{dsfont}
\usepackage{ifthen}
\usepackage[normalem]{ulem}
\newcommand{\bra}[1]{\langle #1 |}
\newcommand{\ket}[1]{| #1\rangle}

\DeclareMathOperator{\Tr}{Tr}

\newcommand{\bmm}{\begin{matrix}}
\newcommand{\emm}{\end{matrix}}
\makeatletter
\newcommand{\thickhline}{%
    \noalign {\ifnum 0=`}\fi \hrule height 1.25pt
    \futurelet \reserved@a \@xhline
}
\newcolumntype{"}{@{\hskip\tabcolsep\vrule width 1.25pt\hskip\tabcolsep}}
\makeatother

\newcommand{\Oop}[1]{
\begin{tikzpicture}
[scale=0.28,baseline={([yshift=-.5ex]current bounding box.center)}]
\path[draw,thick,gray!70] (-1,-1) rectangle (1,1);
\draw (0,0) node {$#1$};
\begin{scope}[rounded corners=12pt,thick,rotate=50,decoration={
    markings,
    mark=at position 0.5 with {\arrow{}}}
    ] 
    \path[draw,postaction={decorate}] (-2,0) -- (0,2) -- (2,0) -- (0,-2) -- cycle;
    \path[draw,postaction={decorate}] (0,-2) -- (-2,0) -- (0,2) -- (2,0) -- cycle;
    \path[draw,postaction={decorate}] (2,0) -- (0,-2) -- (-2,0) -- (0,2) -- cycle;
    \path[draw,postaction={decorate}] (0,2) -- (2,0) -- (0,-2) -- (-2,0) -- cycle;        
\end{scope}

\end{tikzpicture}
}

\newcommand{\OopNumb}[1]{
\begin{tikzpicture}
[scale=0.28,baseline={([yshift=-.5ex]current bounding box.center)}]
\path[draw,thick,gray!70] (-1,-1) rectangle (1,1);
\draw (0,0) node {$#1$};
\draw (-1.3,-1.3) node {\scalebox{0.8}{$4$}};
\draw (-1.3,1.3) node {\scalebox{0.8}{$1$}};
\draw (1.3,-1.3) node {\scalebox{0.8}{$3$}};
\draw (1.3,1.3) node {\scalebox{0.8}{$2$}};

\begin{scope}[rounded corners=12pt,thick,rotate=50,decoration={
    markings,
    mark=at position 0.5 with {\arrow{}}
  }
    ] 
    \path[draw,postaction={decorate}] (-2,0) -- (0,2) -- (2,0) -- (0,-2) -- cycle;
    \path[draw,postaction={decorate}] (0,-2) -- (-2,0) -- (0,2) -- (2,0) -- cycle;
    \path[draw,postaction={decorate}] (2,0) -- (0,-2) -- (-2,0) -- (0,2) -- cycle;
    \path[draw,postaction={decorate}] (0,2) -- (2,0) -- (0,-2) -- (-2,0) -- cycle;        
\end{scope}

\end{tikzpicture}
}

\newcommand{\TwoBlocks}[1]{
\begin{tikzpicture}
[scale=0.5,baseline={([yshift=-.5ex]current bounding box.center)}]
\ifnum#1=1  
        \path[draw,white,fill=gray!20] (0,0) rectangle (0+1,0+1);
        \path[draw,white] (0,1) rectangle (0+1,1+1);
        \draw (.5,.5) node {$\tilde p$};
        \draw (.5,1.5) node {$p$};
        \draw[thick, dotted] (0,-.5) -- (0,2.25);
    \else
        \path[draw,white] (0,0) rectangle (0+1,0+1);
        \path[draw,white,fill=gray!20] (0,1) rectangle (0+1,1+1);
        \draw (.5,.5) node {$p$};
        \draw (.5,1.5) node {$\tilde p$};
        \draw[thick, dotted] (0,-.25) -- (0,2.5);
\fi

\draw[ultra thick] (0,0) -- (0,2);

\end{tikzpicture}
}

\newcommand{\ComEdge}[3]{
\begin{tikzpicture}
[scale=0.22,baseline={([yshift=-.5ex]current bounding box.center)}]
\begin{scope}[rounded corners=12pt,thick,rotate=0,decoration={
    markings,
    mark=at position 0.52 with {\arrow{}}}
    ]
\ifnum#1=1 
    \path[draw,#2,postaction={decorate}] (0,-2) -- (-2,0) -- (0,2);
    \path[draw,white,double=black] (0,-4) -- (-2,-2) -- (0,0);
    \path[draw,#3,postaction={decorate}] (0,-4) -- (-2,-2) -- (0,0);
\else
    \path[draw,#3,postaction={decorate}] (0,-4) -- (-2,-2) -- (0,0);
    \path[draw,white,double=black] (0,-2) -- (-2,0) -- (0,2);
    \path[draw,#2,postaction={decorate}] (0,-2) -- (-2,0) -- (0,2);
\fi
\end{scope}
\end{tikzpicture}
}

\begin{document}

%\preprint{}

\title{Universal edge information from wave-function deformation}


%\author{}
%\thanks{}

\author{Wen Wei Ho}
\affiliation{Department of Theoretical Physics, University of Geneva, 24 quai Ernest-Ansermet, 1211 Geneva, Switzerland}
\affiliation{Perimeter Institute for Theoretical Physics, Waterloo, Ontario, N2L 2Y5 Canada}

\author{Lukasz Cincio}
\affiliation{Perimeter Institute for Theoretical Physics, Waterloo, Ontario, N2L 2Y5 Canada}

\author{Heidar Moradi}
\affiliation{Perimeter Institute for Theoretical Physics, Waterloo, Ontario, N2L 2Y5 Canada}

\author{Guifre Vidal}
\affiliation{Perimeter Institute for Theoretical Physics, Waterloo, Ontario, N2L 2Y5 Canada}

\date{\today}

\begin{abstract}
It is well known that the bulk physics of a topological phase constrains its possible edge physics through the bulk-edge correspondence. Therefore, the different types of edge theories that a topological phase can host constitute a universal piece of data which can be used to characterize topological order. In this paper, we argue that, beginning from only the fixed-point wave function (FPW) of a nonchiral topological phase and by locally deforming it, all possible edge theories can be extracted from its entanglement Hamiltonian (EH).  We give a general argument, and concretely illustrate our claim by deforming the FPW of the Wen-plaquette model, the quantum double of $\mathbb{Z}_2$. In that case, we show that the possible EHs of the deformed FPW reflect the known possible types of edge theories, which are generically gapped, but gapless if translational symmetry is preserved. We stress that our results do not require an underlying Hamiltonian--thus, this lends support to the notion that a topological phase is indeed characterized by only a set of quantum states and can be studied through its FPWs.
% Furthermore, wave-function deformation has practical applications in the context of tensor networks, where FPWs take particularly simple representations.
\end{abstract}

%We consider a wave function perturbation analysis around fixed-point states of systems with topological order. Despite the fact that the entanglement spectrum is flat for fixed-point states, we show that information about edge states can be obtained using our approach. In particular we proof XXXXX using analytical methods and YYYYY by a numerical analysis using a tensor network approach. We carry out this analysis on fixed-point wave functions with $\mathbb Z_2$ topological order, on topologically non-trivial manifolds, and compute the CFT spectrum of the edge for translational invariant perturbations (SET) for each sector and XXXXX. This shit is important because of XXXXXXX.

%\keywords{}

\maketitle

\section{Introduction}

Topological order (TO) in a gapped $(2+1)$-dimensional quantum many-body system is believed to be characterized entirely by universal properties of its ground state(s) \cite{TOGS1,TOGS2, TOGS3, TOGS4}. For instance, a nonzero topological entanglement entropy $\gamma$ in the ground state indicates the presence of TO and is a measure of the total quantum dimension of the underlying anyonic system \cite{TEE1, TOGS4}. The braiding statistics of anyons in the theory is another such universal property and can be extracted from the $\mathcal{S}$ and $\mathcal{T}$ matrices, computed by measuring the overlap between the ground states transformed by modular matrices on a torus \cite{ST1, ST2, ST3, ST4}.

The different kinds of edge theories that a topological phase can support, when placed on a manifold with a boundary, constitute another universal piece of data that we will be concerned with in this paper. It is well-known from the bulk-edge correspondence that the topological physics of the bulk constrains the possible types of edge theories \cite{HalperinEdgeIQHE, WenFQHEEdge92, WenTopOrdersEdge1995, DomainWalls}. For example, in Abelian topological phases, it is understood that the number of topologically distinct gapped edges is in one-to-one correspondence with the number of Lagrangian subgroups of the anyonic model in the bulk, each of which is a set of quasiparticles that obey certain braiding statistics within and without the set \cite{Edge1, Edge2, Edge3}. One very useful way of studying the edge theory of a given bulk is to look at the entanglement spectrum [ES] (or entanglement Hamiltonian [EH]) of its ground state $\ket{\Psi}$, through the edge-ES (or edge-EH) correspondence \cite{ES}. The EH $H_{\text{ent.}}$ is defined as follows.  For a given bipartition of the system into two parts $L$ and $R$, such that the entanglement cut mimics the geometry of the physical edge in question, the EH is obtained from $\rho_L\equiv \frac{1}{Z}e^{-H_{\text{ent.}}}$, where $Z=\Tr(e^{-H_{\text{ent.}}})$ and $\rho_L :=\Tr_R \ket{\Psi}\bra{\Psi}$ is the reduced density matrix on $L$. The ES is then simply the eigenvalues of the EH. The edge-ES correspondence states that the ES typically reproduces the universal, low-energy spectrum of the edge \cite{ESChiral, ESNonChiral}. It is natural to conjecture that this correspondence  applies not just to the spectrum but also to the Hamiltonians in an edge-EH correspondence -- such a view  is indeed supported by the recent work of Ref.~\onlinecite{ESNonChiral}.

However, here a quandary arises. A single quantum state gives a unique EH, i.e. a single instance of an edge theory. Yet, as mentioned before, the universal features of topological order, which include the {\it set} of possible edge theories, should all be contained within a quantum state. Thus, a natural question that arises is: can one extract all possible edge theories starting from only one quantum state (or a microscopically few number of quantum states) believed to host the TO?

In this paper, we argue that this is indeed the case: by locally deforming only one (or a few) quantum state(s), one can extract the edge theories of the TO, at least perturbatively. Concretely, we work with nonchiral topological phases, where the natural quantum states that characterize the TO are the so-called fixed-point wave functions (FPWs), $\ket{\psi_\text{FPW}}$  \cite{TN2, FPW1, LU1, FPW2}. These are special quantum states obtained at the fixed-point of an entanglement renormalization group flow in the space of quantum states, after all nonuniversal, short-ranged entanglement has been removed. We consider deforming the FPW as such:
\begin{align}
\ket{\psi_\text{FPW}} \to \ket{\psi'} \equiv \bigotimes_i (\mathbb{I}_i  + \epsilon V_i) \ket{\psi_\text{FPW}}.
\label{eqn:wfdeformation}
\end{align}
Here $i$ is a  region localized in space (not necessarily single-site), $\epsilon$ a small parameter and $V_i$ some chosen operator with support on $i$. Our claim is that all edge theories of a topological phase, obtained perturbatively from a fixed-point limit (more precisely, the fixed-point Hamiltonian), can be extracted by studying the EH of the deformed FPW $\ket{\psi'}$. Furthermore, we can restrict the set of operators $V_i$ to have support only in $L$, so that we can also study the edge theories through deformations of the reduced density matrix directly: 
\begin{align}
\rho_L \to \rho_L' =  \left[ \bigotimes_{i \in L}  (\mathbb{I}_i  + \epsilon V_i)  \right] \rho_L  \left[ \bigotimes_{i \in L} (\mathbb{I}_i  + \epsilon V_i)  \right].
\label{eqn:rdmdeformation}
\end{align}
Note that there is no bulk Hamiltonian involved in this approach of studying edge theories - thus, wave-function deformation lends even more support to the view that TO is characterized by only a small set of quantum states.

At first sight, the possibility of extracting universal edge information simply from deformations of the FPWs is a rather surprising claim. This is so because the $n$th-R\'enyi entropies of $\rho_L$ are all equal for the FPW, so one would expect that beyond the topological entanglement entropy, no further universal data about the edge can be extracted from $\ket{\psi_\text{FPW}}$, which was indeed claimed in Ref.~\onlinecite{TEERenyi}. On the other hand, TO is characterized by the ``pattern of entanglement'' in the wave function \cite{FPW2}, and so all universal data including that of the edge should be contained within the structure of the FPW.

In the rest of this paper, we present both theoretical analysis and numerical evidence to support the latter point of view. For the sake of exposition, we first illustrate our claim in a specific model:  the Wen-plaquette model \cite{WenPlaq}, a quantum double of $\mathbb{Z}_2$ with TO similar to the toric code. We present a perturbative analysis to derive the EHs of the Wen-plaqutte model from local deformations to its FPWs, comparing this to known results about  its edges theories. We also numerically confirm our analysis by showing that we can reproduce and distinguish the two topologically distinct gapped edge theories that are well known to exist for a system with $\mathbb{Z}_2$-toric code TO, by measuring a nonlocal order parameter. Then, we present a perturbative argument for the validity of wave-function deformation for general nonchiral topological phases that are described by string-net models. Finally, we discuss potential applications of wave-function deformation and conclude.

\section{Example: Wen-plaquette model}
\label{sec:example}

It is instructive to first illustrate concretely our claim of extraction of universal edge information beginning from only the FPW in a specific model, before presenting the argument for general nonchiral topological phases. We will focus on the Wen-plaquettte model in this section.

\subsection{Edge theories of Wen-plaquette model, revisited}
We first review known results about the edge of the Wen-plaquette model using the Hamiltonian approach (mainly following Ref.~\onlinecite{ESNonChiral}; see also Ref.~\onlinecite{Shuo} for a Projected Entangled Pairs States (PEPS) approach).

\begin{figure}
  		\includegraphics[width=.42\textwidth]{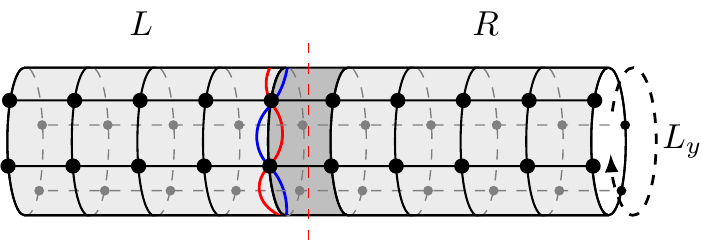}
\caption{(Color online). The infinite cylinder of width $L_y$ on which the Wen-plaquette model is defined on, with the bipartition into two semi-infinite cylinders $L$ and $R$. The red and blue strings acting on the row of spins adjacent to the entanglement cut are the two noncontractible Wilson loops wrapping around the cylinder, $\Gamma^e$ and $\Gamma^m$.}
\label{fig:Cylinder}
\end{figure}

The Wen-plaquette model is a fixed-point Hamiltonian acting on a square lattice of spin-$1/2$s,  comprised of mutually commuting plaquette-terms:
\begin{align}
H = -\sum_{p} \mathcal{O}_p = -\sum_{p}  \Oop{p},
\label{eqn:Hwen}
\end{align}
and its ground states(s) are FPWs. $\mathcal{O}_p =  \scalebox{.8}{\OopNumb{p}}=Z_1X_2Z_3X_4$ is a plaquette-term, where $\{X_i,Y_i,Z_i\}$ are the Pauli-matrices acting on site $i$. The emergent TO is bosonic $\mathbb{Z}_2$-toric code, and so the system supports anyonic quasiparticle excitations labeled by $\{1,e,m,f\}$. The geometry we consider here is an infinite cylinder of circumference $L_y$ ($L_y=4n$ for some integer $n$), with a smooth bipartition dividing the infinite cylinder into two semi-infinite cylinders left ($L$) and right ($R$), mimicking the physical edge of a semi-infinite cylinder.  On such a geometry, there are four topologically distinct FPWs, each of which carries an anyonic flux as measured by the two  noncontractible Wilson loops $\Gamma^e=Z_1X_2\cdots Z_{L_y-1}X_{L_y}$ and $\Gamma^m = X_1Z_2\cdots X_{L_y-1}Z_{L_y}$ wrapping around the cylinder, which we choose to act on the circle of spins on $L$ just adjacent to the entanglement cut, see Fig.~\ref{fig:Cylinder}.

\begin{figure}
  		\includegraphics[width=\columnwidth]{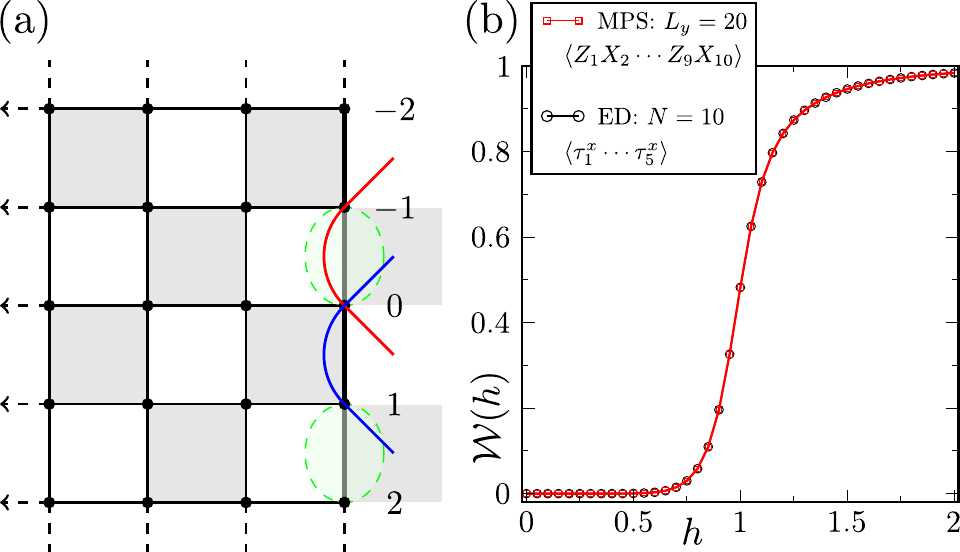}
\caption{(Color online). (a) The $L$ semi-infinite cylinder with the boundary on the right. The numbers label rows of spins. The red boundary operators $Z_{2n-1} X_{2n}$ get mapped to $\tau_n^x$, while the blue boundary operators $Z_{2n}X_{2n+1}$  get mapped to $\tau_n^z \tau_{n+1}^z$. $\tau_n^\alpha$ is an $\alpha$-Pauli operator acting on the emergent DOFs $\tau$ at the edge, a pseudospin-$1/2$, depicted by the green ellipse. (b) $\mathcal{W}(h)$ against $h$ for an infinite cylinder of circumference $L_y=20$ and $\epsilon=0.001$. Red squares represent the numerical results obtained using the ground state of the EH on a semi-infinite cylinder, while black circles represent the exact diagonalization results of a bona fide Ising spin chain of length $N=10$ -- the agreement is virtually perfect. One can clearly see that $\mathcal{W}(h)$ distinguishes between the two phases, ferromagnetic for $h<1$ and paramagnetic for $h>1$, with the critical value at $h=1$.}
\label{fig:LatticeEdge}
\end{figure}

On the left $(L)$ semi-infinite cylinder, which has a boundary, we know from the work of  Ref.~\onlinecite{ESNonChiral} that the emergent degrees of freedom (DOF) which both the edge Hamiltonian (EdH) and the entanglement Hamiltonian (EH) act on are pseudospin-$1/2$s, composed each of two real spins on the boundary (see Fig.~\ref{fig:LatticeEdge}(a)). In addition, the algebra of boundary operators (i.e.~operators which act on these emergent DOFs) is generated by the operators $Z_i X_{i+1}$,  that each act on two boundary spins $(i,i+1)$ of $L$. %One can further label the boundary operators as red or blue depending on where they act: $Z_{2n-1}X_{2n}$ are red boundary operators, while $Z_{2n} X_{2n+1}$ are blue boundary operators (see Fig.~\ref{fig:LatticeEdge}a). Within each set, boundary operators commute, but between sets, operators obey non-trivial commutation relations. 
In terms of the pseudospin-$1/2$s, a mapping to pseudospin operators that preserves the commutation relations of these boundary operators can be found as follows:
\begin{align}
Z_{2n-1} X_{2n} \leftrightarrow \tau_{n}^x, ~~~ Z_{2n} X_{2n+1} \leftrightarrow \tau_{n}^z \tau_{n+1}^z,
\label{eqn:mapping}
\end{align}
where $\tau^\alpha_n$ is an $\alpha$-Pauli operator acting on the $n$-th pseudospin-$1/2$ (see also Fig.~\ref{fig:LatticeEdge}(a)). We see that these are $\mathbb{Z}_2$ symmetric, Ising-type terms $\tau_n^x$ and $\tau_n^z \tau_{n+1}^z$. There is a similar mapping for boundary operators on $R$. Thus, both the EdH and EH are made out of linear combinations of products of $\tau_n^x$ and $\tau_n^z \tau^z_{n+1}$, which are therefore both $\mathbb{Z}_2$ symmetric Hamiltonians.

What are the different gapped edge theories of the Wen-plaquette model?  From the works of Refs.~\onlinecite{Edge1, Edge2, Edge3}, we know that there are two known topologically distinct gapped edges in a system with a bosonic $\mathbb{Z}_2$-toric code TO, which are given by the Lagrangian subgroups $\{1, e\}$ and $\{1,m\}$. In the language of the pseudospin DOFs, $\tau$, and the form of the edge Hamiltonian in terms of boundary operators, these two topologically distinct gapped edge theories can be easily understood as the paramagnetic and ferromagnetic phases of an emergent $\mathbb{Z}_2$ Ising-type Hamiltonian, with the two phases separated by a quantum phase transition described by a $(1+1)$-dimensional, $c = 1/2$ Ising CFT.

Let us now realize a clean, canonical, Ising model on a physical edge to the lowest nontrivial order in perturbation theory. Consider the following perturbation to the bulk Hamiltonian defined on the semi-infinite cylinder $L$, Eq.~(\ref{eqn:Hwen}):
\begin{align}
\epsilon V(h) = -\epsilon \sum_{i \in L} V_i(h), ~ V_i(h) = \begin{cases} Z_i + hX_i, \text{ $i$ even} \\ hZ_i + X_i, \text{ $i$ odd} \end{cases},
\label{eqn:V}
\end{align}
so that the full bulk Hamiltonian is $H+\epsilon V(h)$. Here, $\epsilon\ll 1$ (the bulk gap is $1$), and $h$ is a tunable parameter.  It has been shown in Ref.~\onlinecite{ESNonChiral} that to $\mathcal{O}(\epsilon^2)$, both the EdH on a semi-infinite cylinder $L$ and the EH of the ground state of $H+\epsilon V(h)$ on an infinite cylinder are proportional to (up to a shift) the emergent Ising Hamiltonian:
\begin{align}
H_{\text{Ising}} = - \sum_n \left( \tau^z_n \tau^z_{n+1} + h^2 \tau^x_n \right),
\label{eqn:Hee}
\end{align}
acting on the pseudospin DOFs. The different FPWs (with which to calculate the EdH and EH) give the boundary conditions on a circle (periodic/anti-periodic), and also the different $\mathbb{Z}_2$ symmetry sectors of $G=\prod_n\tau^x_n$ (see Ref.~\onlinecite{ESNonChiral} for a more detailed explanation of the symmetry sectors corresponding to different FPWs).

If $h<1$, the ground state of Eq.~(\ref{eqn:Hee}) realizes the ferromagnetic phase, while if $h>1$, then it realizes the paramagnetic phase. When $h=1$, so that there is full translational symmetry around the cylinder, the EdH and EH are both the critical Ising model, which realizes the $c=1/2$ Ising CFT in the low-energy limit, as expected from Ref.~\onlinecite{ESNonChiral} using arguments of Kramers-Wannier self-duality.

\subsection{Edge theories of Wen-plaquette model from wave-function deformation}
 Our aim now is to recover the phase diagram of Eq.~(\ref{eqn:Hee}) starting from   {\it only} the FPWs of the Wen-plaquette model, and to show how wave-function deformation can be used to extract this information.

To be precise, we work with $\ket{\psi_\text{FPW}}$ that has the identity flux, i.e. it is an eigenstate of both the $\Gamma^e$ and $\Gamma^m$ Wilson loops wrapping around the cylinder with eigenvalues $+1$. This choice of FPW selects for the $\mathbb{Z}_2$-symmetric sector of the Ising Hamiltonian with periodic boundary conditions \cite{BCs}. Now, the FPWs of the model are defined by the flux-free conditions, $\mathcal{O}_p=+1$ for every plaquette $p$. Note that these conditions do not require the notion of a Hamiltonian, even though the states that satisfy these conditions are obviously realized as the ground states of Eq.~(\ref{eqn:Hwen}). The unnormalized FPW  with the identity flux is given by
\begin{align}
\ket{\psi_\text{FPW}} = \prod_{a = e,m} \left( \frac{\mathbb{I} + \Gamma^a}{2}\right)\prod_p \left(\frac{\mathbb{I}+\mathcal{O}_p}{2} \right) \ket{0}_L\ket{0}_R,
\label{eqn:FPW}
\end{align}
where $\ket{0}_L \ket{0}_R$ is a reference state.  $\frac{1}{2}(\mathbb{I}+\mathcal{O}_p)$ is a projector onto the flux-free sector, and $\prod_{a = e,m}\frac{1}{2}(\mathbb{I}+\Gamma^a)$ projects onto the $+1$ eigenvalues of the noncontractible Wilson loops on the cylinder, $\Gamma^e$ and $\Gamma^m$.
\begin{comment}
 that the deformed FPW becomes
\begin{align}
&\ket{\psi'(h)} = \sum_{\tau} \left(\mathbb{P}_{+1} - 2\epsilon^2 \mathbb{P}_{+1} H_{\text{Ising}}^{\text{PBC}}\right) \ket{\tau}_L \ket{\tau}_R + \cdots,
\end{align}

, where $\cdots$ refer to terms such as $\mathcal{O}(\epsilon) \ket{\alpha}_L \ket{\tau}_R$. 

we see the $\cdots$ terms do not contribute to the entanglement Hamiltonian at leading order, and so
\end{comment}

Now, to recover the edge physics of the Wen-plaquette model, we locally deform only the $L$ half of the FPW given by Eq.~(\ref{eqn:wfdeformation}), using $V_i=V_i(h)$ as in Eq.~(\ref{eqn:V}), so that $\ket{\psi_\text{FPW}}\to\ket{\psi'(h)}$. We find, combining (1) a perturbative calculation in the representation of the FPW in terms of pseudospin variables $\tau$ (see the Appendix A for the details), and (2) the detailed calculations performed in Appendices B and C of Ref.~\onlinecite{ESNonChiral},  that
\begin{align}
\rho_L' &= \mathcal{N'} \exp\left(-4 \epsilon^2 \mathbb{P}_{+1} H_{\text{Ising}}^{\text{PBC}}\right) + \mathcal{O}(\epsilon^3),
\label{eqn:deformedRho}
\end{align}
restricted to the $G=+1$ symmetry sector. That is, $H_{\text{ent.}}$ is proportional to the $\mathbb{Z}_2$-symmetric periodic Ising model, Eq.~(\ref{eqn:Hee}), which is also proportional to the edge Hamiltonian. In contrast, for the (undeformed) FPW,
\begin{align}
\rho_L &= \mathcal{N} \mathbb{P}_{+1}.
\end{align}
Eq.~(\ref{eqn:deformedRho}) is an instance of the concrete expression of our claim -- that the EH $H_\text{ent.}$ in $\rho_L'$, obtained only from deformations to the wave function of the Wen-plaquette model, indeed informs us about the edge physics, given by Eq.~(\ref{eqn:Hee}). Note the striking contrast between $\rho_L$ of the FPW and  $\rho_L'$ of the deformed FPW: the former has a flat ES and only tells us about the topological entanglement entropy of the topological phase, while the latter has an ES that gives us information about the edge.

However, if the EH of the deformed FPW not only reproduces the spectrum of the EdH, but is also proportional to it, then we should be able to directly obtain the phase diagram of Eq.~(\ref{eqn:Hee}) by measuring a suitable order parameter in the ground state $\ket{\text{GS}(h)}_L$ of the EH. Typically, the order parameter that distinguishes between the ferromagnetic and paramagnetic phases in the Ising Hamiltonian is the local order parameter $\tau^z_n$, which detects symmetry breaking. However, because Eq.~(\ref{eqn:Hee}) is actually an emergent Hamiltonian acting on pseudospin DOFs, certain emergent operators cannot be realized by the underlying, original, degrees of freedom. In particular, there is no way to realize the $\mathbb{Z}_2$-odd local operator $\tau^z_n$ (which one would typically measure to detect symmetry breaking in such a model) in terms of the local boundary operators $Z_iX_{i+1}$, as the latter all get mapped to $\mathbb{Z}_2$-even operators (see Eq.~(\ref{eqn:mapping})). One therefore has to measure a nonlocal order parameter to distinguish between the two phases; two possible choices are the {\it open} string operators
\begin{align}
&W^e = Z_1 X_2 \cdots Z_{L_y/2-1} X_{L_y/2} \leftrightarrow  \tau_1^x \tau_2^x \cdots \tau_{L_y/4}^x, \nonumber \\
&W^m = Z_2 X_3 \cdots Z_{L_y/2} X_{L_y/2 + 1} \leftrightarrow \tau_1^z \tau_{L_y/4}^z,
\label{eqn:OrderParameter}
\end{align}
acting on spins of $L$ adjacent to the entanglement cut (they are {\it not} the closed string operators $\Gamma^e$ or $\Gamma^m$). Intuitively, $W^e$ and $W^m$ measure the amount of anyonic condensation of $e$ and $m$ quasiparticles respectively on the boundary \cite{Edge1, Edge2}. Since these two operators are Kramers-Wannier duals of each other, it suffices to measure only one; we choose to measure only $W^e$. The expectation value is then computed in the ground state of the EH:
\begin{align}
\mathcal{W}(h) := \bra{\text{GS}(h)}_L W^e \ket{\text{GS}(h)}_L.
\end{align}
When $L_y \to \infty$, the order parameter should show a kink at $h = 1$ where the quantum phase transition is. For $h < 1$, $\mathcal{W}(h)$ should be vanishing, signifying the ferromagnetic phase, while for $h > 1$, $\mathcal{W}(h)$ should increase as a power law $\mathcal{W}(h) \sim (h-1)^\beta$ with some critical exponent $\beta$, and saturate at $+1$, signifying the paramagnetic phase. 

\begin{comment}
\begin{figure}
\center
\includegraphics[width=0.45\textwidth]{W_h.pdf}
\caption{(Color online). $\mathcal{W}(h)$ against $h$ for an infinite cylinder of circumference $L_y=20$ and $\epsilon=0.001$. Red squares represent the numerical results obtained using the ground state of the EH on a semi-infinite cylinder, while black circles represent the exact diagonalization results of a bona fide Ising spin chain of length $N=10$ -- the agreement is virtually perfect. One can clearly see that $\mathcal{W}(h)$ distingiushes between the two phases, ferromagnetic for $h<1$ and paramagnetic for $h>1$, with the critical value at $h=1$.}
\label{fig:OrderParameter}
\end{figure}
\end{comment}

We implement this procedure and obtain the phase diagram numerically. We utilize an exact representation of $\ket{\psi_\text{FPW}}$ on an infinite cylinder with circumference $L_y$, encoded in a matrix product state (MPS) that wraps around the cylinder in a snake-like fashion \cite{ST2}. We deform the MPS according to Eq.~(\ref{eqn:wfdeformation}) with deformations given by Eq.~(\ref{eqn:V}), and then extract the Schmidt vector corresponding to the largest singular value in the Schmidt decomposition, which gives us $\ket{\text{GS}(h)}_L$.

Fig.~\ref{fig:LatticeEdge}(b) shows the plot of $\mathcal{W}(h)$ against the tuning parameter $h$, for $L_y = 20$ and $\epsilon=0.001$ (the results are insensitive to the exact values of $\epsilon$ as long as $\epsilon\ll 1$). As expected, $\mathcal{W}(h)$ shows a sudden increase from $0$ in the region $h<1$ to $+1$ in the region $h>1$, with the transition at $h=1$. For comparison we have also plotted $\mathcal{W}(h)$ of a bona fide Ising spin chain of length $N=L_y/2=10$ with periodic boundary conditions, Eq.~(\ref{eqn:Hee}), obtained via exact diagonalization. The agreement is virtually perfect. This shows that we have successfully extracted the two known gapped edges in this system with $\mathbb{Z}_2$-toric code TO, by locally deforming only $\ket{\psi_\text{FPW}}$. Note crucially that at no stage of the numerical illustration was there any optimization of the MPS tensors.

\begin{comment}
 For the Wen-plaquette model, this is done as follows. Let $\ket{v_0(h)}_L$ denote the ground state of $\rho_L'$, which equivalently is the Schmidt vector $\ket{v_0(h)}_L$ corresponding to the largest Schmidt value in the decomposition
\begin{align}
\ket{\psi'(h)} = \sum_n \lambda_n(h) \ket{v_n(h)}_L \otimes \ket{v_n(h)}_R.
\end{align}
$\ket{v_0(h)}_L$ is then also the ground state of the emergent Ising Hamiltonian $H_{\text{Ising}}^{\text{PBC}}$ of Eq.~(\ref{eqn:Hee}), acting on the $L$ semi-infinite cylinder.

%; this is why we could obtain the phase diagram as shown in Fig.~\ref{fig:OrderParameter}. solely from wave-function deformation.
\end{comment}

%%%%%%%%%%%%%

\section{General argument for nonchiral topological phases}

Having illustrated how wave-function deformation works in a concrete example, the Wen-plaquette model, we now make the case for the validity of wave-function deformation in general nonchiral topological phases. As seen from the preceding section, the crucial point was that both the EdH of the Wen-plaquette model and the EH obtained through just a local deformation of its FPW act on the same emergent DOFs and are generated by the same algebra of boundary operators. Thus, wave-function deformation of the FPW could be used to explore the space of edge theories and extract the desired universal edge information, which we did successfully.

Our aim in this section is to therefore argue that the same line of reasoning is true for general nonchiral topological phases. Concretely, we consider nonchiral topological phases for which there is a string-net description, and perform a Schrieffer-Wolff(SW) transformation\cite{SchriefferWolff} to derive the edge Hamiltonian and entanglement Hamiltonian, beginning from a fixed-point Hamiltonian and its corresponding fixed-point wave function respectively. We will see that the forms of both the edge Hamiltonians of a given theory and the entanglement Hamiltonians of its deformed FPWs (given by Eqs.~(\ref{eqn:wfdeformation}) or (\ref{eqn:rdmdeformation})) are the same, both being generated by the same algebra of boundary operators acting on the same emergent DOF. Thus, this would imply that the universal edge information of a topological phase is contained within the FPW and can be extracted by locally deforming the FPW and studying its EH. Note that since the arguments presented in this section are perturbative in nature, they do not constitute a mathematical proof of our claim; however, the calculation done for the specific case of the Wen-plaquette model in the previous section, together with the numerical evidence presented, constitute strong evidence for the validity of the line of reasoning below.

\subsection{Edge Hamiltonian}
 The starting point is the fixed-point Hamiltonian of a generic nonchiral topological phase on a lattice, described by a string-net model\cite{FPW2}:
\begin{align}
H = -\sum_{i \subset \mathcal{M}} P_i,
\label{eqn:FPH}
\end{align}
where $P_i$ are commuting projectors, acting on a spatially local (not necessarily single-site) region $i$ of a closed manifold $\mathcal{M}$, so that the ground state subspace consists of states that satisfy $P_i = +1$ for all $i$. The ground state subspace is typically degenerate, split into different topological sectors, differentiated by noncontractible Wilson loops around the manifold.

Note that the condition $P_i = +1$ for all $i$ also precisely defines the fixed-point wave functions of the nonchiral topological phase, so that a Hamiltonian is actually not needed to describe these wave functions which characterizes this phase. However, we will use this fixed-point Hamiltonian to give meaning to the term `edge theories' of a topological phase.

Consider now if the manifold $\mathcal{M}$ is instead open, so that it has a boundary $\partial \mathcal{M}$. Then, if Eq.~(\ref{eqn:FPH}) still describes the Hamiltonian on $\mathcal{M}$, in addition to the microscopic ground state degeneracy given by the different topological sectors, there will be a macroscopic ground state degeneracy within each topological sector, given by emergent local degrees of freedom (DOF) on the boundary. One can remove this degeneracy by imposing boundary conditions on $\partial \mathcal{M}$, which amounts to adding small $(\epsilon \ll 1)$ {\it local} perturbations $\epsilon V$ acting near $\partial \mathcal{M}$, so that the full Hamiltonian is $H + \epsilon V$. The edge theory of the topological phase  can then be understood as the low energy subspace of $H+\epsilon V$, which can be calculated, for example, using the Schrieffer-Wolff(SW) transformation:
\begin{align}
H_{\text{edge}} = \mathbb{P}_0 H \mathbb{P}_0 + \epsilon \mathbb{P}_0 V \mathbb{P}_0 + \frac{1}{2} \epsilon^2 \sum_{j \neq 0} \frac{\mathbb{P}_0 V \mathbb{P}_{j}V \mathbb{P}_0}{E_j - E_0} +\cdots,
\label{eqn:SW}
\end{align}
where $\mathbb{P}_0$ is the projector onto the ground state subspace, and $\mathbb{P}_j$ is the projector onto the higher energy subspaces. We see from the above expansion that the only terms which contribute to the edge Hamiltonian are those that commute with all $P_i$, since they preserve the ground state condition $P_i = +1$. That is, $[\mathbb{P}_0 V \mathbb{P}_0, P_i]=0$, $[\mathbb{P}_0 V \mathbb{P}_{j \neq 0} V \mathbb{P}_0, P_i] = 0$, and so on. 

We define $\mathcal{A}$ as the maximal set  of algebraically independent Hermitian operators that each acts locally on the manifold $\mathcal{M}$ and which commutes with all $P_i$. That is, $\mathcal{A}$ is comprised of algebraically independent operators $a_j$ which satisfy
\begin{align}
[a_j,P_i] = 0 \text{ for all } i,
\end{align}
such that support($a_j$) is in $\mathcal{M}$. Obviously, all $P_i$ and the noncontractible Wilson loops are in $\mathcal{A}$, but on an open manifold, there will be typically many more local operators $a_j$, with support near the boundary $\partial \mathcal{M}$, localized around site $j$, that also satisfy this condition. We will hence call them `boundary operators'. The boundary operators of $\mathcal{A}$   then generates  (by virtue of being a maximal set of algebraically independent operators) the edge Hamiltonian  -- that is, $H_\text{edge}$, given by Eq.~(\ref{eqn:SW}), must be a linear combination of products of $a_j$:
\begin{align}
H_\text{edge} = \sum c_{i_1,\cdots, i_l}  a_{i_1}\cdots a_{i_l},
\end{align}
where $c_{i_1,\cdots,i_l}$ is a   coefficient denoting the weight of the string of boundary operators $a_{i_1}\cdots a_{i_l}$ and the sum is over such strings.

If the  discussion  above seems cryptic, it is instructive to go back to the example of the Wen-plaquette model considered in the previous section. There, the Hamiltonian Eq.~(\ref{eqn:FPH}) is given by Eq.~(\ref{eqn:Hwen}), and an example of a boundary operator $a_i \in \mathcal{A}$ is $Z_i X_{i+1}$, which we see is both localized near the boundary and commutes with all plaquette operators $\mathcal{O}_p$ in the bulk.  Furthermore, the set of all such boundary operators $Z_i X_{i+1}$  indeed generates  $H_\text{edge}$, see Eqs.~(\ref{eqn:mapping}) and (\ref{eqn:Hee})~(see also Ref.~\onlinecite{ESNonChiral}).

Note that the boundary operators $a_i$ in general obey nontrivial commutation relations between themselves, which give rise to an algebra $\mathcal{B_A}$ that we will call the `boundary operator algebra' (c.f. Eq.~(\ref{eqn:mapping}) for the Wen-plaquette model). Because of this nontrivial algebra, the edge theory will have a nontrivial dispersion relation. Also, if the operators $a_i$ are chosen as local as possible, then the condition that one adds {\it local} perturbations $\epsilon V$ to $H$ translates to the fact that the edge Hamiltonian will also obey some approximate sense of locality on $\partial \mathcal{M}$, since the support of the terms in the expansion of Eq.~(\ref{eqn:SW}) can only grow linearly with the order of $\epsilon$. Also, a local perturbation $V$ ensures that the edge theory can be defined within each topological sector of the bulk theory without ambiguity, as mixing between topological sectors will be suppressed by an exponentially small factor $\sim \epsilon^L$, where $L$ is the length scale associated with the boundary $\partial M$.

\subsection{Entanglement Hamiltonian of deformed FPW}

Now, we shift perspectives and start from the fixed-point wave functions of a topological phase defined  on a closed manifold $\mathcal{M}$. We assume an entanglement cut of $\mathcal{M}$ into two parts $L$ and $R$, mimicking the physical cut. Our aim in the following is to argue that the entanglement Hamiltonian that emerges from wave-function deformation of the FPW is also generated by $\mathcal{A}$, the maximal set of algebraically independent operators that act locally and which commutes with all $P_i$. If so, then that would imply that the space of EHs and the space of EdHs are equivalent (at least perturbatively), and so edge information of the topological phase can be extracted from wave-function deformation, thereby supporting our claim.

We have the following Schmidt decomposition of the FPW:
\begin{align}
\ket{\psi_\text{FPW}} = \Gamma \prod_{i \cap \partial M \neq 0} P_i \ket{P_{j \subset L} = +1} \ket{P_{k \subset R} = +1},
\end{align}
where $P_{j/k}$ are projectors that have support entirely in $L/R$, while projectors $P_i$ are those that span the entanglement cut. $\Gamma$ is some Wilson line/loop that chooses the FPW of a particular topological sector. Since $[P_i,P_{j/k}]=0$, it must be that $P_i$ can be decomposed into products of elements of $\mathcal{A}_L$ and $\mathcal{A}_R$. Here $\mathcal{A}_\xi$ is the set $\mathcal{A}$, defined previously, corresponding to the $\xi = L/R$ semi-infinite cylinder. That is, schematically, $P_i = \sum_\mu f^\mu_L(a_L) f^\mu_R(a_R)$, for some functions $f^\mu_{L/R}$. As an example, a plaquette term in the Wen-plaquette model that straddles the entanglement cut is $Z_{L,1} X_{R,1} Z_{R,2} X_{L,2}$ which can be written as $Z_{L,1} X_{L,2} \otimes X_{R,1} Z_{R,2}$, where the two terms of the tensor product belong to $\mathcal{A}_L$ to $\mathcal{A}_R$ respectively.

Thus, the Schmidt decomposition of the FPW must be
\begin{align}
\ket{\psi_\text{FPW}} = \frac{1}{\sqrt{N}}\sum_{\mu=1}^{N} \ket{a^\mu_L}\otimes \ket{a^\mu_R},
\end{align}
where $N$ is the multiplicity that gives the correct topological entanglement entropy of the topological phase, and $\ket{a^\mu_{L/R}}$ are states in the ground state subspace of the {\it open} manifolds $\mathcal{M}_{L/R}$, schematically distinguished by the boundary operators $a^\mu_{L/R} \in \mathcal{A}_{L/R}$ (recall the macroscopic degeneracy in the case of open manifolds which can be resolved by the boundary operators). This gives the reduced density matrix on $L$:
\begin{align}
\rho_L = \frac{1}{N} \mathbb{P}_0^\Gamma,
\label{eqn:fprdm}
\end{align}
where $\mathbb{P}_0^\Gamma$ is the projector onto the ground state manifold (of $H$ as in Eq.~(\ref{eqn:FPH})) restricted to the topological sector chosen by $\Gamma$. We see from this that the ES is flat, as the ES of the FPW should be.

Next we deform the reduced density matrix as in Eq.~(\ref{eqn:rdmdeformation}). If we write $\bigotimes_i (\mathbb{I}_i + \epsilon V_i) \approx \bigotimes_i \exp(\epsilon V_i)  \approx \exp({\sum_{n \geq 1} \epsilon^n S_n})$ ($S_n$ is given by the Baker-Campbell-Hauserdoff formula; in particular, $S_1 = \sum_i V_i$), then we have the deformed reduced density matrix
\begin{align}
\rho'_L \approx \frac{1}{N} \left(\underbrace{\mathbb{P}_0^\Gamma}_{h} + \underbrace{ \epsilon\{ S_1, \mathbb{P}_0^\Gamma \}  + \mathcal{O}(\epsilon^2)}_{v} \right),
\label{eqn:drdm}
\end{align}
where $\{ \cdot \}$ is the anticommuator and we have interpret Eq.~(\ref{eqn:drdm}) as a `perturbation' $v$ on a `unperturbed' Hamiltonian $h$. We now apply the SW transformation to the Hamiltonian $h+v$, like before, but  to instead obtain the `higher-energy' subspace perturbatively. Here $||h|| =1$ and $||v|| \sim \epsilon$ so the use of the SW transformation is justified. We have, to the two lowest orders in $\epsilon$,
\begin{align}
(\rho'_L)_\text{high} = \frac{1}{N}\left( \mathbb{P}_0^\Gamma + \epsilon \mathbb{P}_0^\Gamma v \mathbb{P}_0^\Gamma + \frac{\epsilon^2}{2} \frac{\mathbb{P}_0^\Gamma v \mathbb{P}_1^\Gamma v \mathbb{P}_0^\Gamma}{1-0} \right).
\label{eqn:SWrdm}
\end{align}
Comparing the above to the perturbative expansion of the edge Hamiltonian given by Eq.~(\ref{eqn:SW}), there is more than structural similarity of the expansions; there is also {\it algebraic} similarity. Since $\mathbb{P}_0^\Gamma$ is non other than the projector onto the ground state manifold of the fixed-point Hamiltonian $H$ on $\mathcal{M}$, it follows that $(\rho'_L)_\text{high}$ must also be generated by $\mathcal{A}$, the maximal set of local, algebraically independent operators in $\mathcal{M}$ that commute with all $P_i$, similar to $H_\text{edge}$. Furthermore, since the deformations of the FPW were local to begin with, $(\rho'_L)_\text{high}$ is also approximately local on $\partial \mathcal{M}$. There is one more step to the entanglement Hamiltonian $H_\text{ent.}$: one has to take the logarithm of $(\rho'_L)_\text{high}$, but it is clear that the entanglement Hamiltonian will also be generated by $\mathcal{A}$, although the notion of locality might be affected. However, to lowest order in $\epsilon$, the entanglement Hamiltonian will still  be approximately local. An explicit example of the above calculations for a particular model, the Wen-plaquette model, can be found in Appendix A and also Ref.~\onlinecite{ESNonChiral}.

Thus, we have argued that the forms of both the edge and entanglement Hamiltonians are the same, both being generated by $\mathcal{A}$. This implies that the space of entanglement Hamiltonians obtained from a local deformation of the FPW is the same as the space of edge Hamiltonians obtained from various bulks, at least perturbatively. It is natural to assume then that for some local perturbation $V$ to the {\it fixed-point Hamiltonian} $H$ which gives some edge Hamiltonian, there exists a suitable choice of local deformations $\{ V_i \}$ to the {\it FPW} such that the EH reproduces the edge Hamiltonian. Of course, our arguments above do not provide an explicit recipe for constructing this map; this map depends on the specific nonchiral topological phase in question, as one would have to find both the set of operators $\mathcal{A}$ and also the boundary operator algebra $\mathcal{B_A}$ that these operators satisfy. However, we believe that we have managed to present convincing arguments for our claim: that universal edge information can be extracted solely from local deformations of the FPWs of a nonchiral topological phase.

\begin{comment}
 The expansion of the deformation $\bigotimes_i (\mathbb{I}_i + \epsilon V_i)$, to each order in $\epsilon$, will have terms that are generated solely by $\mathcal{A}_L \otimes \mathcal{A}_R$, and terms that are not spanned by this algebra. Then, typically, the most general unnormalized state after the deformation, to lowest order, is
\begin{align}
\ket{\psi'} &= \sum_{\mu,\nu} (\delta_{\mu,\nu} + \epsilon H_{\mu,\nu}) \ket{a_L^\mu} \otimes \ket{a_R^\nu} \nonumber \\
& + \epsilon \sum_{\mu,\nu} C_{\mu,\nu} \ket{a_L^\mu} \otimes \ket{\neq a_R^\nu} + D_{\mu,\nu} \ket{\neq a_L^\mu} \otimes \ket{a_R^\nu} \nonumber \\
& + \epsilon \sum_{\mu,\nu} E_{\mu,\nu} \ket{\neq a_L^\mu} \otimes \ket{\neq a_R^\nu}.
\end{align}
Here,

\end{comment}

\section{Discussion and conclusion}
In this paper, we have argued through both analytical and numerical means that using wave-function deformation on the FPWs, one can extract the different edge theories that a nonchiral topological phases can support, at least perturbatively.  We stress that this process does not require a bulk Hamiltonian, as firstly the FPW can be defined by local consistency relations, and secondly the deformation is done at the wave function level. Since the different edge theories that a topological phase can support is a universal piece of data of the TO, this lends support to the belief that TO is characterized solely by a set of quantum states. 

wave-function deformation can potentially be used to distinguish between systems with different TO. For example, two FPWs can have the same topological entanglement entropy (such as the $\mathbb{Z}_2$ Kitaev toric code and $\mathbb{Z}_2$ double semion which both have $\gamma = \log 2$), but extraction of the different edge theories they can host can be used to further differentiate between them. Furthermore, the study of edge theories using wave-function deformation can be readily applied to other nonchiral topological phases, especially since FPWs take simple representations in terms of tensor networks \cite{TN1, TN2, TN3, TN4} -- in particular, this allows for a numerically relatively inexpensive way of exploring the space of edge theories as no numerical optimization is indeed. For instance, a study of the edge theories of the $\mathbb{Z}_3$ Wen-plaquette model has been conducted \cite{unpub}. 

As a closing remark, we note that the analysis done in this paper was perturbative in nature, controlled by the small parameter $\epsilon$. Since we see that we can go from the FPW to any gapped or gapless boundary type, and since the local deformation is invertible, it follows that we can go from any boundary type to any boundary type of the topological phase, starting from a perturbative deformation of the FPW. This is likely to be true also for any nonperturbative deformation, as long as we do not destroy the TO. However, here one would potentially have to `dress' the order parameter operators (Eq.~(\ref{eqn:OrderParameter})) appropriately, see Ref.~\onlinecite{LU1}. It may thus be possible to explore the entire phase diagram of edge theories of a topological phase starting from a state $\ket{\psi}$ with a certain edge theory (i.e. not necessarily the FPW): one could move in this phase space of edge theories by locally deforming $\ket{\psi}$ (nonperturbatively) to produce another state $\ket{\psi'}$ with a different edge theory, even if the two edge theories are separated by a phase transition.

\textit{Acknowledgments.} -- L.C. and G.V. acknowledge support by the John Templeton Foundation. G.V. also acknowledges support by the Simons Foundation (Many Electron Collaboration). This research was supported in part by Perimeter Institute for Theoretical Physics. Research at Perimeter Institute is supported by the Government of Canada through Industry Canada and by the Province of Ontario through the Ministry of Research and Innovation.

\bibliography{refs}

\begin{thebibliography}{32}
\bibitem [{BCs()}]{BCs}%
  \BibitemOpen
  \href@noop {} {}\bibinfo {note} {If we had chosen to work with the other
  FPWs, then we get entanglement Hamiltonians which correspond to Ising models
  with different boundary conditions and symmetry sectors. Namely, the FPWs
  with the $e$ or $m$ anyonic flux give rise to entanglement Hamiltonians that
  are the periodic Ising model in the $G=-1$ symmetry sector or the
  antiperiodic Ising model in the $G=+1$ symmetry sector (the spectra of both
  are the same), while the FPW with the $f$ anyonic flux gives rise to an
  entanglement Hamiltonian which is the antiperiodic Ising model in the $G=-1$
  symmetry sector.}\BibitemShut {Stop}%
\end{thebibliography}

%\begin{comment}
% \newpage

\begin{appendix}
\section*{Appendix: Perturbative calculation of entanglement Hamiltonian (EH)}
\label{sec:Appendix}

First we rewrite the FPW of the Wen-plaquette model, Eq.~(\ref{eqn:FPW}), in terms of boundary pseudospin-$1/2$ degrees of freedom, $\tau$, as explained in the main text and in Ref.~\onlinecite{ESNonChiral}. This representation will also illustrate the pattern of entanglement ($\mathbb{Z}_2$-toric code TO) contained in the wave function.

The product over the plaquettes $p$ in Eq.~(\ref{eqn:FPW}) splits into $3$ sets: those that act on $L$, those that act on $R$, and those that act on the strip of spins where the entanglement cut is defined through. Define $\ket{L}$ as $\prod_{p \in L} \frac{1}{2}(\mathbb{I} + \mathcal{O}_p) \ket{0}_{L}$ and similarly for $R$. Here $\ket{0}_L \ket{0}_R$, the reference state in Eq.~(\ref{eqn:FPW}), is chosen in such a way that $( Z_{2n} X_{2n+1})_L \ket{L}\ket{R} = (X_{2n} Z_{2n+1})_R \ket{L}\ket{R} = \ket{L}\ket{R}$ for all $n$, where $n$ labels the spins on both $L$ and $R$ adjacent to the entanglement cut (i.e. this fixes the gauge of the reference state).

With this choice of reference state, $\ket{L}$ can be represented as the state with pseudospin configuration $\ket{\uparrow \uparrow \cdots\uparrow \uparrow}_L$ (i.e. all $\tau_n$s are pointing up), and there is a similar representation for $\ket{R}$. Furthermore, the mapping of boundary operators (e.g.~$Z_i X_i$ acting on $L$) to pseudospin operators is given by Eq.~(\ref{eqn:mapping}). Since the plaquettes acting on the strip (through which the entanglement cut is made) are comprised of a product of two boundary operators from the $L$ and $R$ cylinders, the FPW can be written as a superposition of pseudospin configurations on the $L$ and $R$ halves:
\begin{align}
\ket{\psi_\text{FPW}} = \sum_{\tau} \mathbb{P}_{+1}\ket{\tau}_L \ket{\tau}_R = \sum_{\tau} \ket{\tau_+}_L \ket{\tau}_R,
\label{eqn:FPW2}
\end{align}
where $\mathbb{P}_{+1}$ is the projector on the $G = \prod_n \tau_n^x = +1$ symmetry sector, and $\ket{\tau}$ is a state with a certain pseudospin configuration (e.g. $\ket{\uparrow \downarrow \downarrow\cdots \downarrow \uparrow}$). Two different pseudospin configurations are orthogonal: $\langle \tau' | \tau \rangle = \delta_{\tau',\tau}$, and $\ket{\tau_+} = \ket{\tau} + \ket{\bar{\tau}}$, where $\bar{\tau}$ is the completely flipped configuration of $\tau$. Ignoring the projector, one can intuitively see that this state is a loop quantum gas -- it is an equal weight superposition of loops on the cylinder. The different configurations $\tau$ correspond to the different ways loops cross the entanglement cut; $\ket{\tau}_L$ must pair with only $\ket{\tau}_R$ or $\ket{\bar{\tau}}_R$ in order to form a closed loop.

We deform the FPW $\ket{\psi_\text{FPW}}$ of the Wen-plaquette model, Eq.~(\ref{eqn:FPW}) (or Eq.~(\ref{eqn:FPW2})), according to Eq.~(\ref{eqn:wfdeformation}) with $V_i = V_i(h)$ as given by Eq.~(\ref{eqn:V}), and calculate the EH of the reduced density matrix $\rho_L$. Note that the manipulations here are formally similar to that of Ref.~\onlinecite{ESNonChiral}, but the logic is fundamentally different: there, the perturbative calculation was performed for deformations to the Hamiltonian, while here, the perturbative calculation is performed for deformations to the wave function.

Now, we note that the $V_i$s split into two sets -- those that act on spins in the bulk of $L$ (that is, away from the entanglement cut), and those that act on the circle of spins in $L$ living adjacent to the entanglement cut. The former set simply renormalizes $\ket{\tau}_L \to \ket{\tilde{\tau}}_L$, which is still an orthogonal set, and so we drop the tilde label in our discussion. We therefore see that the change of the entanglement spectrum comes only from deformations to the wave function on spins next to the entanglement cut.

The deformed FPW, to $\mathcal{O}(\epsilon^2)$, is then
\begin{align}
\ket{\psi'(h)} &= \prod_{i \in L, \text{adj. to cut}} (\mathbb{I}_i + \epsilon V_i) \sum_\tau \ket{\tau_+}_L \ket{\tau}_R \nonumber\\
& = \left( \mathbb{I} + \epsilon \sum_i V_i + 2 \epsilon^2 \sum_{i <j} V_i V_j  \right) \sum_\tau \ket{\tau_+}_L \ket{\tau}_R.
\label{eqn:expansion}
\end{align}
Consider the $\mathcal{O}(\epsilon)$ effect of the deformation. This generates terms $\ket{\alpha_+}_L \ket{\tau}_R$ where $\ket{\alpha}_L$ is a new ket orthogonal to all the pseudospin configurations $\ket{\tau}_L$ (specifically it is a state describing an excitation in the bulk). Consider next the $\mathcal{O}(\epsilon^2)$  effect of the deformation. This generates two kinds of states. If $V_i$ and $V_j$ are not adjacent, then we also obtain a state $\ket{\alpha_+}_L \ket{\tau}_R$. But if $j = i+1$ i.e. that $V_i$ is next to $V_{j}$, then they can form boundary operators $Z_i X_{i+1}$, so that the deformed FPW contains new states $\ket{\tau'}_L \ket{\tau}_R$ for some $\tau', \tau$. The crucial point is that there is now additional coupling between states that are labeled only by pseudospin configurations which are beyond the diagonal $\ket{\tau_+}_L \ket{\tau}_R$ ones. These off-diagonal terms $\ket{\tau'}_L\ket{\tau}_R$ generate the EH. 

Specifically, from the mapping given by Eq.~(\ref{eqn:mapping}), $V_{2n-1} V_{2n} \ket{\tau_+}_L \ket{\tau}_R \leftrightarrow h^2 \tau_n^x \ket{\tau_+}_L \ket{\tau}_R + \cdots$ and $V_{2n} V_{2n+1} \ket{\tau_+}_L \ket{\tau}_R + \leftrightarrow \tau_n^z \tau_{n+1}^z \ket{\tau_+}_L \ket{\tau}_R + \cdots$, so that the deformed FPW is 
\begin{align}
&\ket{\psi'(h)} \nonumber \\
&=  \sum_{\tau} \left(\mathbb{I} + 2\epsilon^2 \sum_n (\tau^z_n \tau^z_{n+1} + h^2 \tau^x_n)  \right) \ket{\tau_+}_L \ket{\tau}_R + \cdots \nonumber \\
& = \sum_{\tau} \left(\mathbb{P}_{+1} - 2\epsilon^2 \mathbb{P}_{+1} H_{\text{Ising}}^{\text{PBC}}\right) \ket{\tau}_L \ket{\tau}_R + \cdots,
\end{align}
where $\cdots$ refer to terms such as $\mathcal{O}(\epsilon) \ket{\alpha}_L \ket{\tau}_R$. At this stage, we are done: from the detailed calculation performed in Appendices B and C of Ref.~\onlinecite{ESNonChiral}, we see the $\cdots$ terms do not contribute to the EH at leading order, and so
\begin{align}
\rho_L' &= \Tr_R \ket{\psi'(h)} \bra{\psi'(h)} \nonumber \\
& = \mathcal{N}' \exp\left(-4 \epsilon^2 \mathbb{P}_{+1} H_{\text{Ising}}^{\text{PBC}}\right) + \mathcal{O}(\epsilon^3).
\end{align}
That is, the EH is proportional (up to a constant shift) to the periodic Ising Hamiltonian projected into the $G = \prod_n \tau_n^x = +1$ sector, which in turn is proportional to the edge Hamiltonian (EdH) of the Wen-plaquette model. This is Eq.~(\ref{eqn:deformedRho}) in the main text.

\end{appendix}
%\end{comment}

\end{document}